**Estimating IRI based on pavement distress type, density, and severity: Insights from machine learning techniques**

**Julie Yu Qiao**
Graduate Research Assistant, Center for Connected and Automated Transportation (CCAT), and Lyles School of Civil Engineering, Purdue University, West Lafayette, IN, 47907.
Email: qaio14@purdue.edu

**Sikai Chen**
Visiting Assistant Professor, Center for Connected and Automated Transportation (CCAT), and Lyles School of Civil Engineering, Purdue University, West Lafayette, IN, 47907.
Email: chen1670@purdue.edu; and
Visiting Research Fellow, Robotics Institute, School of Computer Science, Carnegie Mellon University, Pittsburgh, PA, 15213.
Email: sikaichen@cmu.edu
ORCiD: 0000-0002-5931-5619

**Majed Alinizzi**
Assistant Professor, Department of Civil Engineering, College of Engineering, Qassim University, 51452 Buraydah, Qassim, Saudi Arabia
Email: alinizzi@qec.edu.sa

**Miltos Alamaniotis**
Assistant Professor, Electrical & Computer Engineering Department, University of Texas-San Antonio, San Antonio, TX, 78249, USA
Email: miltos.alamaniotis@utsa.edu

**Samuel Labi**
Professor, Center for Connected and Automated Transportation (CCAT), and Lyles School of Civil Engineering, Purdue University, West Lafayette, IN, 47907.
Email: labi@purdue.edu
ORCiD: 0000-0001-9830-2071

**ABSTRACT**
Surface roughness is primary measure of pavement performance that has been associated with ride quality and vehicle operating costs. Of all the surface roughness indicators, the International Roughness Index (IRI) is the most widely used. However, it is costly to measure IRI, and for this reason, certain road classes are excluded from IRI measurements at a network level. Higher levels of distresses are generally associated with higher roughness. However, for a given roughness level, pavement data typically exhibits a great deal of variability in the distress types, density, and severity. It is hypothesized that it is feasible to estimate the IRI of a pavement section given its distress types and their respective densities and severities. To investigate this hypothesis, this paper uses data from in-service pavements and machine learning methods to ascertain the extent to which IRI can be predicted given a set of pavement attributes. The results suggest that machine learning can be used reliably to estimate IRI based on the measured distress types and their respective densities and severities. The analysis also showed that IRI estimated this way depends on the pavement type and functional class. The paper also includes an exploratory section that addresses the reverse situation, that is, estimating the probability of pavement distress type distribution and occurrence severity/extent based on a given roughness level.

**INTRODUCTION AND BACKGROUND**

**Background**
A key aspect of pavement management systems is the needs for reliable pavement condition data because this data are used to assess pavement network condition, to schedule maintenance and rehabilitation (M&R), and to estimate the level of funding needed for M&R. Reliable data quality leads to adequate characterization of pavement condition, right timing of M&R investments, reliable reporting of the effectiveness of individual projects or systemwide programs, and ultimately, avoidance of wasteful spending of agency resources or user frustration due to unduly deferred maintenance.

At the current time, the need for reliable data is exacerbated by resource constraints at state highway agencies. In their quest for cost-effective techniques and to minimize human error and resource consumption that are associated with manual techniques, agencies have resorted to automated data collection techniques (Wang and Gong 2005; Zalama et al. 2014). Such automation has been facilitated by advancements in computer and information technology, particularly for capturing and processing digital images (Wang 2011; Chen et al., 2019; Chen et al., 2020). Fully automated methods involve minimal human involvement in data collection process. Typically, the operator drives the sensor-equipped vehicle driven over the pavement section and the pavement condition information is collected at normal driving speed. The pavement condition is often reported in terms of a performance indicator (PI), which are standard measurements that reflect the extent and the severity of surface distresses and defects including cracking, rut depth, and roughness.

The need to report pavement condition for a wide variety of distress indicators, across the entire carriageway width, and over several miles means that massive amounts of data are collected. Therefore, automated methods represent both virtue and vice: while facilitating data collection and reduction of human effort, they produce enormous amounts of data that make it difficult to analyze using manual or semi-manual techniques. If the relationship between these indicators can be quantified with a high degree of confidence, then there is little or no need to collect data on all the distress indicators: data on a single indicator could be used to predict the occurrence or severity of other indicators. If that is the case, then the data collection and processing effort (time and cost) can be drastically reduced. This paper addresses this issue in the specific context of surface roughness.

**Surface Roughness**
Surface roughness can be defined as the bumpiness of a ride due to significant deviations from a truly smooth surface thus affecting vehicle dynamics, ride quality, dynamics loads, and drainage (Vedula et al. 2003). This is an important pavement characteristic because it affects the dynamics of vehicle movement, leads to increasing wear of vehicle parts, and therefore has significant impacts on vehicle operating costs (Islam and Buttler, 2014), travel speed (Wang et al, 2014), fuel consumption (Zaabar and Chatti, 2010), ride comfort ride and safety (Molenaar and Sweere, 1978). As we discuss in a subsequent section of this paper, past research suggests that pavement roughness is related to surface condition (Park, 2007) and to disaggregate distress types (Al-Mansour and Alawal, 2006). Given the increasing availability of pavement distress data arising from the automation of monitoring processes at agencies worldwide, there is increased opportunity to carry out a deeper investigation of the relationships between aggregate performance and individual distresses.

The International Roughness Index (IRI), is the most common indicator of pavement surface roughness. The use of IRI is popular because it generally represents the user perception of road quality and it is widely used by many highway agencies to serve as a basis for network-level performance monitoring and pavement repair decisions. IRI is often reported in units of meters per kilometer or inches per mile. However, it is costly to measure IRI, and for this reason, certain road classes are excluded from IRI measurements at a network level. For road sections that have no IRI measurements, it is possible that knowledge of the surface distresses can help provide an estimate of the surface roughness within a fair degree of accuracy.

Intuitively, a higher level of pavement distress attributes (types, extent, severity) is generally associated with a higher level of roughness. For example, a pavement that exhibit extensive and severe potholes, faults, and bumps, is likely to exhibit a high level of roughness. What is needed is a numerical relationship accompanied by the degree of confidence in such a relationship, and the factors that affect the strength of the relationship. This is because the relationship is not always strong, for two reasons. First, certain distress types do not translate into roughness until in the long term where they metamorphose into other distress types that are more strongly linked to surface roughness. For example, extensive occurrence of cracking may not immediately reflect as surface roughness because as they travel over the crack, the vehicle tires bridge over these cracks without causing a bump to the vehicle. Secondly, for a given roughness level, there typically exhibits a great deal of variability in the distress types, density, and severity. For example, two pavements that have the same roughness level may be having very different distress types and the distributions, extent, and severity of these distress types.

In light of the foregoing discussion, it can be hypothesized that within some bounds of confidence, it may be feasible to estimate the IRI of a pavement section on the basis of its distress types and their respective densities and severities. The question will be: how reliable is this relationship? Is the relationship strength influenced by the pavement surface material type or functional class? Which analytical tools can best bring out these relationships? To investigate these hypotheses, this paper uses data from in-service pavements in a Midwestern state of the USA and three machine learning methods – Support Vector Machine (SVM), logistic regression and Naïve Bayes.

**Pavement Distresses and Performance Indicator**
The factors that influence the appearance of pavement distress types and their occurrence attributes (extent and severity), include traffic loading (Ahmed et al., 2015), climate (Jeong et al, 2017), layer material type and subgrade quality (Vaillancourt et al., 2014), construction quality (Brock and Hedderich, 2007), drainage (Hall and Crovetti, 2007), and external factors such as utility cuts (Alinizzi et al., 2018). Of the various pavement distresses, cracking is a major concern because if not repaired, it allows moisture infiltration of the pavement leading to other distresses and increased surface roughness in the long term (Owusu-Ababio, 1998). Cracks are caused by factors that include traffic loading, temperature changes, water ingress or high water table, or post-frost heave. There are various types of cracks that occur on both flexible and rigid pavements, and those considered in this paper include longitudinal, alligator, block, transverse, edge and corner cracking (SHRP, 2000). Rutting, as a primary criterion of pavement structural performance, is generally defined as a longitudinal surface depression along the wheel paths of pavements, resulting from deformation or consolidation of any of the pavement layers and subgrade. This can be due to heavy traffic loading, insufficient pavement thickness or compaction, and poorly-designed asphalt mixtures, or a combination of these factors. Rutting is a serious issue for road users, because there is an increasing potential for hydroplaning when water accumulates in the ruts and in certain regions, freezes in wintry conditions (Archilla, 2000).

Pavement roughness is a measure of the irregularities in the pavement surface that have adverse effects on ride quality. Roughness is typically reported using a widely-accepted indicator, the International Roughness Index (IRI), originally developed by Gillespie (1994) and adopted by the World Bank in the 1980s (Sayers, 1986). IRI, a standardized roughness measurement used to characterize the longitudinal profile of a traveled wheel-track, is measured as the ratio of the accumulated suspension motion (in mm, inches, etc.) to the distance traveled by a standard vehicle (km, mi, etc.) (Pantha,2010). The units of IRI are millimeters per meter (mm/m), meters per kilometer (m/km), or inch per mile (in/mi) (Zhou, 2008). The use of IRI has increased over the years and that indicator is now a universal indicator of pavement performance (Meegoda, 2014) as it is used in various strategic and operational decision contexts at the network and project levels respectively. As such, IRI is measured regularly by most highway agencies worldwide (Greene et al., 2013). The universal appeal of pavement roughness in general (and IRI specifically) as a performance indicator is not only due to but also the effect of several factors and situations. First, it is used by researchers and highway agencies as the basis for evaluating pavement life-cycle benefits (Lu, 2012), prioritizing pavement maintenance and rehabilitation programs, network-level budget allocation for pavement repair (Haider, 2011) and analysis of tradeoffs across the various management systems (Bai et al., 2014; Ghahari et al., 2019; 2020). Second, pavement roughness manifests as ride bumpiness which profoundly affects ride comfort (Karan, 1976), the main concern of road users. Third, pavement roughness is directly related to vehicle operating cost through adversities associated with a bumpy pavement, such as increased fuel consumption (e.g., Archondo-Callao and Faiz 1994) and vehicle depreciation and tire cost (Haugodegard, 1994). Fourth, pavement roughness, to some extent, influences road safety (Saleh, 2017; Labi et al., 2017; Chen et al., 2017; 2019; Chen T. et al., 2020; Tang et al., 2018; Chen 2019). Finally, rougher roads can decrease the efficiency of a vehicle, increasing fuel use and greenhouse gas emissions (Greene et al., 2013).

**Relationships between IRI and other Indicators of Performance and Distress**
Pavement roughness has been long believed to be correlated to aggregate and disaggregate performance indicators, and researchers have investigated the relationships between IRI and other performance indicators and distresses using empirical data. Al-Omari and Darter (1995) found that IRI increases in a nearly linear fashion with increasing number of transverse cracks, potholes, depressions, and swells. Mactutis et al. (2000) developed linear regression models of IRI as a function of cracking and rut severity. Similarly, Al-Mansour and Alawal (2006) investigated the correlations between visual inspections measurements and surface roughness by developing a linear IRI regression model as a function of four distress types: cracking, patching, number of depression, and raveling. Hozayen and Alrukaibi (2009) developed several regression models to investigate the correlations between pavement roughness on one hand, and rutting, raveling and cracking on the other hand, and found a significant relationship between pavement roughness and raveling distress. They explored a variety of functional forms including linear, exponential, and polynomial, and found that the polynomial form provided the best $R^2$ value. Gharaibeh et al. (2009) assessed the agreement among various indices of pavement condition. Also, Prasad et al. (2013) studied the relationships between roughness and individual pavement distresses.

**Modeling Techniques Used to Develop the Relationships**
Most of the studies that investigated the relationships between IRI and other indicators of overall performance or distresses used linear or nonlinear regression. A few used more complex model structures such as artificial neural networks (ANN). However, so far, the obtained relationships have been characterized by significant uncertainty partly attributed to the subjective nature of the performance or condition ratings and evidenced by low goodness-of-fit of the developed relationships. This situation, obviously, is likely due to lack of network-level data on pavement performance and distress levels. Fortunately, as highway agencies increasingly deploy increasingly sophisticated equipment that measure roughness and multiple other distress types automatically, quickly, and objectively (Adey, 2017), it has become increasingly feasible to carry out a comprehensive analysis of the said relationships using reliable data.

The use of machine learning and deep learning methods such as artificial neural networks (ANNs) in developing pavement distress relationship models was explored by Kaseko and Ritchie (1993) presented a methodology for automating the processing of highway pavement video images using an integration of artificial neural network models with conventional image processing techniques and Koutsopoulos et al. (1993) discussed methods for classifying pavement distress images. Dougherty (1995) summarized the findings of research papers that addressed the application of neural networks in transportation. Mariani et al. (2012) used a normalized truncated Levy walk to describe pavement deterioration. A number of researchers have done beyond using machine learning for describing pavement surface condition; they have used it for prescription of pavement treatments (Zhou and Wang, 2012; Li et al., 2014). Other machine learning applications in relevant literature include Coifman et al., 1998; Chan et al., 2001; Yuan and Cheu, 2003; Wilson et al., 2006; Markovic et al., 2015; Gallego et al. 2018; Lasisi and Attoh-Okine, 2018.

Lin et al. (2003) applied neural networks to investigate the correlations between IRI and pavement distresses including rutting, alligator cracking, potholes, patching, bleeding, corrugation, and stripping. They found that severe potholes, patching, and rutting have the highest correlation to IRI and concluded that there is a significant improvement (drop in roughness) subsequent to treatments that address various distress types. Chandra (2012) developed empirical linear/nonlinear regression models and ANNs models between road roughness and five distress parameters including potholes, patchwork, rut depth, raveling, and cracking using data from Indian highways. He found that the performance of the ANN model was superior to that of the regression based models. The study also proposed an algorithm that allows a comparison of the relative importance (percentage) of the explanatory variables included in the ANNs model. The result shows that the variables like potholes (23%), total cracking (23%), and raveling (20%) are more significant than the patch work (19%) and rut depth (15%) in estimating pavement roughness. From the literature review, it can be hypothesized that machine learning techniques can be used to enhance further, the identification of reliable relationships between surface roughness and pavement distresses. To investigate this hypothesis, this paper uses three commonly-used Machine Learning models (Support Vector Machine, Logistic Regression and Naïve Bayes) to develop predictive models for IRI based on various distress parameters.

## METHODOLOGY
In this paper, machine learning methods were chosen over traditional regression analysis due to the former's superior capacity to handle high dimensional data. Three commonly-used methods were tested and their performance compared.

### Support Vector Machine
The support vector machine (SVM) is a supervised machine learning algorithm that analyzes data based on statistical learning theory that was introduced by Friedman et al. (2001). In general, if the data (with different labels) can be perfectly separated using a hyperplane, the SVM can help establish the maximal margin hyperplane (the hyperplane that is farthest from the training observations) (James et al., 2013). Consider the construction of a maximal margin hyperplane based on $n$ training observations: $x_1, x_2 \ldots x_n \epsilon R^p$ and with corresponding labels $y_1, y_2 \ldots y_n \epsilon \{-1, 1\}$, the maximal margin hyperplane solution can be established using optimization techniques (Eqn 1).

$$Maximize\ M(\beta_0, \beta_1 \ldots \beta_p, \varepsilon_1, \ldots \varepsilon_n)$$
(Eq. 1)
$$s.t.\ \sum_{j=1}^{p} \beta_j = 1$$
$$y_i(\beta_0 + \beta_1 x_{i1} + \beta_2 x_{i2} + \cdots + \beta_p x_{ip}) \geq M(1 - \varepsilon_i) \quad \forall i = 1, \ldots, n$$
$$\varepsilon_i \geq 0,\ \forall i = 1, \ldots, n \qquad \sum_{i=1}^{n} \varepsilon_i \leq C$$

where C is a non-negative constant, M is the width of the margin, $\varepsilon_1, \ldots \varepsilon_n$ are slack variables that allows individual observations to stay in the "wrong" side of the margin.

The "one-against-one" approach (Knerr et al., 1990) was used for multi-class SVM classification. The developed SVM classifiers predict the IRI in a certain range. For data that are not linearly separable, the kernel function $K(x, x_i)$ can be used to replace the hyperplane function which transforms the input space into a high-dimensional space using a nonlinear transformation defined by an inner product function. In this paper, two popular kernel functions (radial basis function or BRF (Eqn 2) and polynomial kernel functions (Eqn 3)) were applied and compared.

BRF Kernel Function: $K(x, x') = exp(-\gamma ||x - x'||^2)$ (Eq. 2)
Polynomial Kernel Function: $K(x, x') = (x^T \cdot x' + 1)^d$ (Eq. 3)

**Logistic Regression**
In machine learning, logistic regression is a probabilistic classifier, borrowed from the field of statistics. In general, a logistic regression model is used to predict the probabilities of the possible outcomes of a dependent variable, given a set of independent variables (Hosmer, 2013). In the case where the dependent variable has more than two outcomes, the term of multinomial logistic regression is used. The general equation for multinomial logistic regression is given in Eqn 4. In this paper, the dependent variable is the IRI of a pavement segment, and the independent variables are a set of pavement distress parameters and other factors such as the road functional class. The multinomial logistic regression is used to predict the probability that the IRI is within a certain range or class *k*, and each range is treated as a discrete outcome.

$$P(C_k) = \frac{e^{\beta_{0,k}+\beta_{1,k}X_{1,k}+\beta_{2,k}X_{2,k}\cdots+\beta_{n_k,k}X_{n_k,K}}}{\sum_{i=1}^{N} e^{\beta_{0,i}+\beta_{1,i}X_{1,i}+\beta_{2,k}X_{2,k}\cdots+\beta_{n_i,i}X_{n_i,i}}}$$ (Eq. 4)

where, $P(C_k)$ indicates the probability of the dependent variable falls in class K ($C_k$); $\beta_0$ is the model constant and the $\beta_{0,1}, \ldots, \beta_{n_k,k}$ are the unknown parameters that correspond to the independent variables ($X_k \text{ where } k = 1, \ldots, n_k$); $n_k$ is the number of independent variables included for class k, and $N$ is the total number of classes. These unknown parameters are typically estimated using maximum likelihood methods.

**Naïve Bayes Classifier**
The Naïve Bayes classifier, another simple probabilistic classifiers, is based on Bayes' theorem and a naive assumption that the features are conditionally independent. While this assumption is not necessarily true, it nevertheless simplifies the estimation dramatically and yields results that work well and often outperform far more sophisticated alternatives (Domingos and Pazzani 1997). A naive Bayes classifier combines the simple probability model with a decision rule. In this paper, the maximum a-posteriori (MAP) decision rule is used. In other words, we used the hypothesis that is most probable. The Bayes classifier with this decision rule is a function (Eqn 5) that assigns a class label $C_k$ (IRI = *k* in this paper) to observation *y*.

$$\hat{y} = argmax \, P(C_k) \prod_{i=1}^{n} P(x_i|C_k)$$ (Eq. 5)

Where ŷ is the estimator for *y*; $P(C_k)$ is the probability that *y* belongs to class *k*; and $P(x_i|C_k)$ is the conditional probability of observing $x_i$ given the condition that *y* belong to class *k* (likelihood).

**DATA DESCRIPTION**

**Data Source**
The data used in this paper's analysis were collected by a vendor contracted by a highway agency in Midwestern USA. This involved automated collection and processing of the pavement surface condition data and video images using an instrumented vehicle equipped with high-resolution cameras and sensors. The instrumented vehicle is driven on one lane at a time (for multi-lane roads, the far-right lane is often monitored because that lane is more likely to experience high truck traffic loading and thus becomes a higher priority for condition assessment. The vehicle surveys the pavement surface using standard methods to report the pavement's overall performance (roughness) and the severity and extent of individual distress types including cracking, rutting and faulting, at regular intervals such as 0.005 driven miles. The data were collected during the physical years 2012-2014 at 50,400 road segments on three roads (I-70, US-52 and US-41). The data were collected at pavements on these three road corridors only. Nevertheless, the number of observations is very large, relatively larger than datasets of past similar efforts – 50,400 pavement sections – due to high level of disaggregation (average condition for every 0.005-mile pavement segment) and the large number of distress variables. Table 1 presents the characteristics of the overall pavement performance (IRI) and distress types (rutting, faulting, and cracking). The abbreviation WP means wheel path.

The distress types studied are: cracking (non-wheel path, wheel path, edge, transverse, block, spall, shoulder, corner), rutting, and faulting. For the crack distresses, the cracking types included alligator, longitudinal, transverse, or a combination of these. The measurement variables for each distress type included length(ft), width(in), depth(in), percentage (%), number, or a combination of these, and the severity levels were: low, medium, and high.

**Data Statistics**
This section presents the basic statistics of the characteristics of the overall pavement performance (IRI) and distress types (rutting, faulting, and cracking). Regarding the cracking distress, the data includes the following occurrence attributes: length or percentage, depth, width, and severity (low, medium, high). A preliminary examination of the data showed that the non-wheel path (Non-WP) Longitudinal Crack, Transverse Crack, Edge Longitudinal Crack and wheel path (WP) Longitudinal Crack are the most common crack types in the dataset. The data distribution of IRI, rutting, faulting and the length, width, and depth of the most common crack type (non-wheel path longitudinal crack) showed that most of the data are normally or with zero or slight skew. In the data, the IRI generally ranges between 30 to 300in/mile, and rutting ranges mostly between 0.02 to 0.2 inches. Faulting has a nearly-normal distribution with most observations located between 0.25 to 0.25. Regarding the cracking attributes, the histogram plot suggests a generally-decreasing frequency as crack length increases, for each of the three severity levels. For each level of crack severity, the crack width and crack depth was found to exhibit a somewhat skewed normal distribution with different means across the different severity types. For example, regarding WP Longitudinal Crack, the most frequent widths at low, medium and high severity levels are 0.21 in, 0.44 in and 0.52 in, respectively. Similar patterns were also found for most of the other crack types.

**MODELING RESULTS AND DISCUSSIONS**
The three machine learning methods (naïve Bayes classifier, SVM and logistic regression) were applied to the dataset. 58 variables were selected in the featureset including road functional class, pavement surface material type (asphalt, concrete), pavement distresses such as faulting, rutting, and so on) and for the various crack types, the relevant attributes (length, width, depth). To validate and compare the proposed models, 80% of randomly selected data were used to train the classifiers and to tune the model parameters. The remaining 20% was used as a test dataset, that is, to evaluate the models and to compare the model performance across different machine learning methods.

In this section of the paper, we present the results using the naïve bayes classifier, and those of the support vector machine (SVM) and logistic regression. Then we compare the results across the three machine learning techniques and interpret the contribution of distress parameters to IRI.

**Results Using The Naïve Bayes Classifier**
Figure 2 presents the testing results for the naïve Bayes classifier for a randomly selected 100 observaitons from the testing data (the remaining observations of the testing data provided similar results and hence were excluded from this figure for brevity). The blue curve represents the actual

IRI values of these observations, while the green curve represents the predicted IRI values using the developed naïve Bayes classifier.

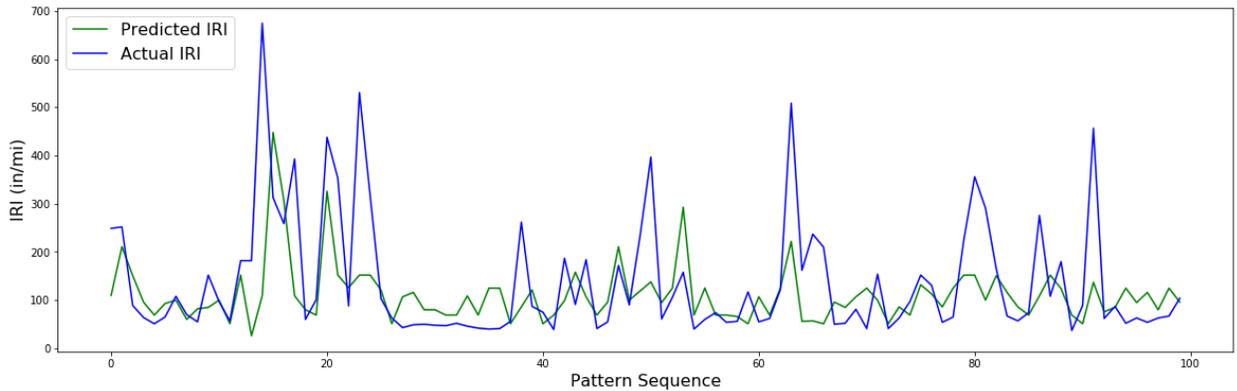

Figure 2. Testing results for Naïve Bayes classifier (100 randomly selected observations)

The data used in this paper is highly disaggregate in the sense that the length for each road segment (observation) is only 0.005 mile. In some cases, this level of granularity is beneficial, particularly where the purpose of the analysis is to evaluate pavement distresses that do not continuously span a lengthy road section or distresses that exist only over relatively short stretches. In such cases, using disaggregate data generally provides more accurate results compared to analysis using aggregate data. In such highly disaggregate datasets, the observed IRI values can be extremely high at a very short pavement segment, for example, an isolated pothole or bridge approaches. It can be observed from the Figure 2 that the devloped model does not predict the IRI values accurately at areas where the actual IRI observed is extremely high (e.g. IRI > 300) This situation exemplifies the realization that in its excess, granularity can be a vice rather than a virtue. To address this issue, we aggregated the data such that each new observation represents a 0.1-mile road segment. Then the naïve Bayes classifier was trained again using the aggregate data. The testing result of using aggregate data is shown in Figure 3 (a) for 30 randomly-selected data points. It was noticed that the prediction accuracy of the model for 0.1-mile segments was significantly higher compared to the initial model (that which used 0.005-mile segments). In addition, due to the existence of noise and error in the data, removing the outlier could further enhance the model performance. As shown in the IRI histogram (Figure 1(a)), only a small percentage of pavement segments have "very high" roughness (IRI > 300 in/mile) and these segments were treated as outliers in this paper and therefore excluded from the analysis. Figure 3(b) presents the results of the models trained on the disaggregate data and aggregate data (after outlier removal).

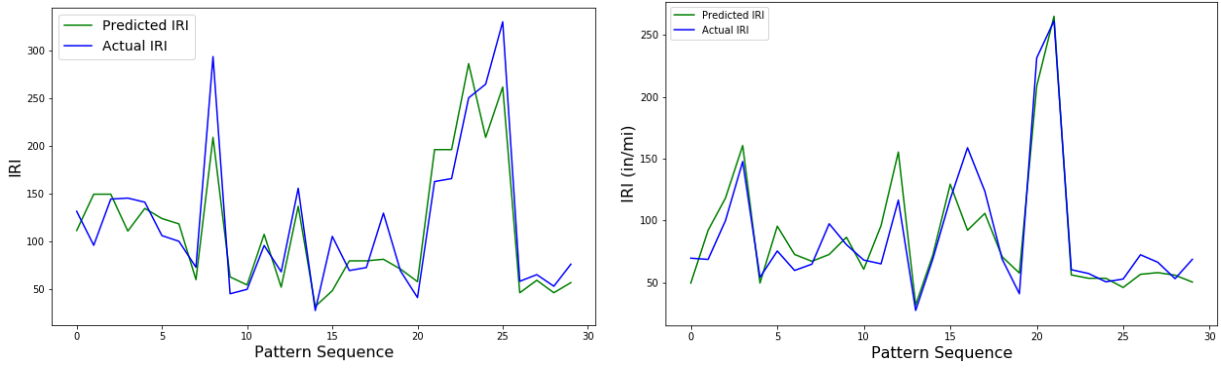

(a) Aggregated Data           (b) Aggregated Data (with outliers removed)
Figure 3. Testing Results for naïve Bayes classifier (30 randomly-selected observations)

Regarding the models developed using naïve Bayes classifiers trained with each of the four categories of processed training data, their performance were compared in terms of prediction accuracy (AC) under different error tolerances. The prediction accuracy under a given error tolerance is estimated using Eqn 6:

$$AC_{toler=T} = \frac{N_{|predIRI-actIRI<T|}}{N_{Test}}$$
(Eq. 6)

Where: $AC_{toler=T}$ is the prediction accuracy under an error tolerance $T$, $N_{Test}$ is the total number of observations in the testing data, $predIRI$ is the predicted IRI and $actIRI$ is the actual IRI, $N_{|predIRI-actIRI|<T}$ is number of observations in the testing data that meet the criteria: the absolute difference between the predicted IRI and the actual IRI is less than $T$.

Table 1 and Figure 4 summarize the performance of the four models: disaggregate data with and without outliers, aggregate data with and without outliers. A significant improvement was found on the prediction accuracy due to the data aggregation and the removal of outliers. After the two processes, the prediction accuracy increased by 13% under the tight error tolerance scenario (prediction error smaller than 20 in/mile) and 19% under the relaxed error tolerance scenario (prediction error smaller than 50 in/mile).

Table 1 Performance comparison of the developed models

|  | Error Tolerance (in/mile) | | |
| --- | --- | --- | --- |
|  | 20 | 30 | 50 |
| **Models with different training data** | | | |
|  | Prediction Accuracy | | |
| Original Disaggregate Data | 0.57 | 0.67 | 0.78 |
| Disaggregate Data (outliers removed) | 0.59 | 0.70 | 0.83 |
| Aggregate Data | 0.68 | 0.77 | 0.92 |
| Aggregate Data (outliers removed) | 0.72 | 0.80 | 0.93 |
|  | | | |
| **SVM models with different kernels** | | | |
|  | Prediction Accuracy | | |
| SVM (RBF Kernel) | 0.78 | 0.84 | 0.96 |
| SVM (Polynomial Kernel) | 0.84 | 0.86 | 0.96 |

| Logistic Regression | | 0.67 | 0.80 | 0.91 |
| --- | --- | --- | --- | --- |
| | | | | |
| **Comparison across the three machine learning techniques** | | | | |
| | | Prediction Accuracy | | |
| Naïve Bayes Classifier | | 0.72 | 0.80 | 0.93 |
| SVM Model (Polynomial Kernel) | | 0.84 | 0.86 | 0.96 |
| Logistic Regression Model | | 0.67 | 0.80 | 0.91 |

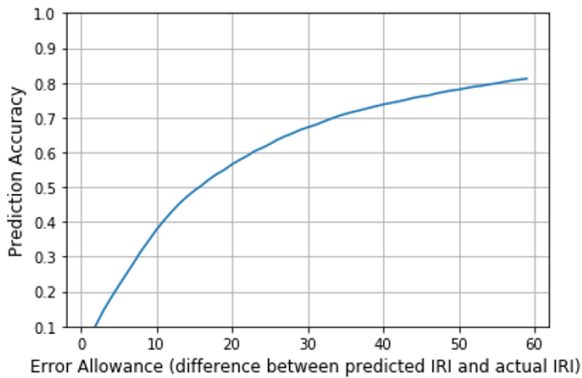
(a) Original disaggregate data

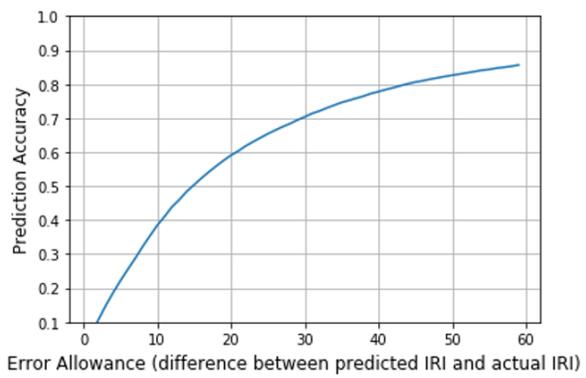
(b) Disaggregate data (with outliers removed)

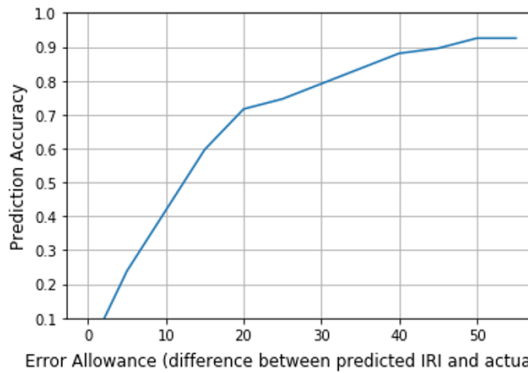
(c) Aggregate data

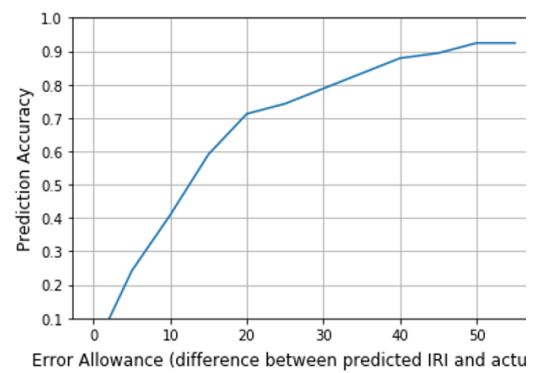
(d) Aggregate data (outliers removed)

Figure 4 Performance comparison between models with four different sets of training data

**Result of Support Vector Machine (SVM) and Logistic Regression**
The SVM and logistic regression were then applied to the data that are aggregated to 0.1 mile per segment and with outliers removed. Table 4 and Figure 5 present the results of the model performance for different kernel functions, and Figure 6 compares their prediction accuracies with different kernels. It was found that the SVM with polynomial kernel has a slightly superior performance compared to RBF kernel. Figure 7 presents the performance of the models.

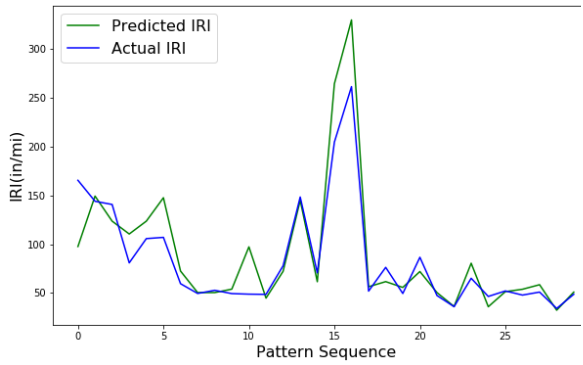
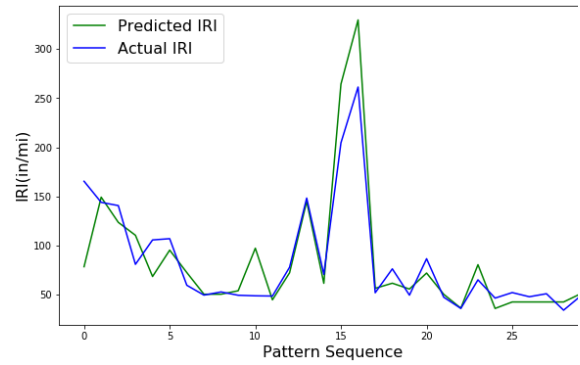

(a) RBF Kernel  (b) Polynomial Kernel

Figure 5 Test Results for SVM models with different kernels
(30 randomly-selected observations)

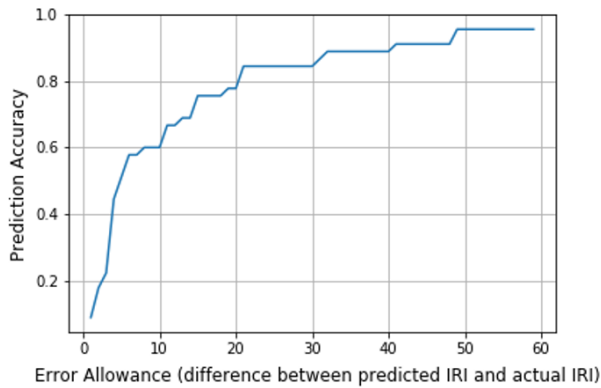
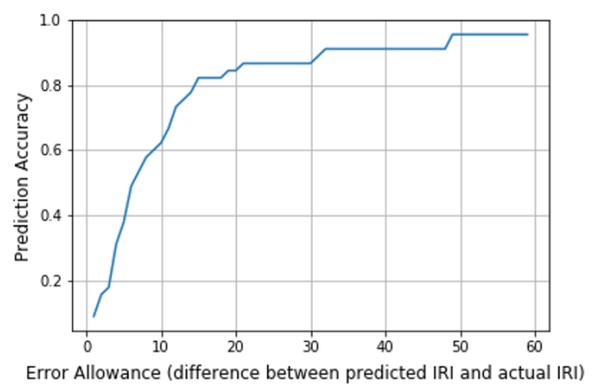

(a) RBF Kernel  (b) Polynomial Kernel

Figure 6 Performance Comparison for SVM models with different kernels
(under different error tolerances)

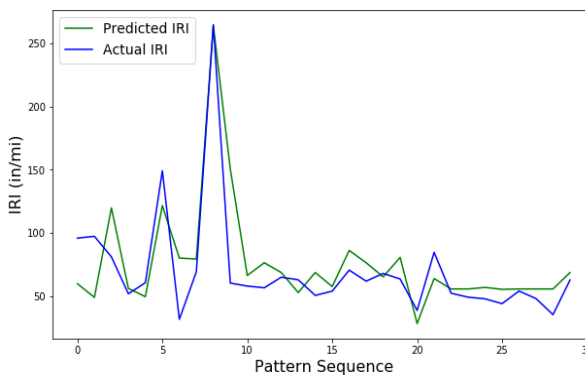
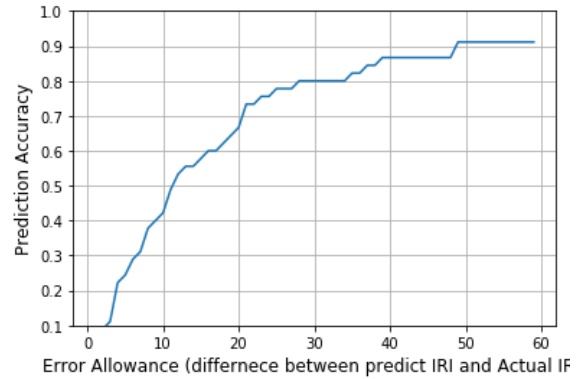

(a) Testing results (30 randomly selected observations)  (b) Model performance (under different error tolerances)

Figure 7 Model results of logistic regression

**Comparison across Different Machine Learning Techniques**
Using the testing data, the performance of the developed models were compared based on their prediction accuracy. From the results (summarized in Table 5 and Figure 8), it was found that the classifiers trained using naïve Bayes and logistic regression methods have similar performance, and that using SVM method in general has the best performance in terms of prediction accuracy. When the error tolerance is set at 20 in/mile, the prediction accuracies for the naïve Bayes classifier, SVM with polynomial kernel and logistic regression were found to be 0.72, 0.84 and 0.67, respectively; for a tolerance of 30 in/mile, the prediction accuracies increase to 0.80, 0.86 and 0.80, respectively. As the tolerance relaxes further to 50 in/mile, the prediction accuracies for these models increase to 0.93, 0.96 and 0.91. In our analysis, the SVM classifier was found to yield superior accuracy; in general, however, all the models indicated that the IRI of a pavement can be predicted reliably given the pavement distress types exhibited by a pavement section and their occurrence attributes.

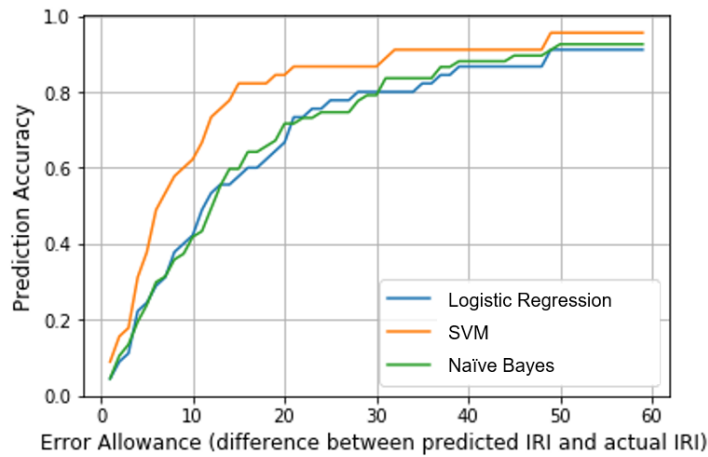

Figure 8 Performance comparison for different machine learning techniques
(under different error tolerances)

**Interpretation of the Contribution of Distress Parameters to IRI**
Machine learning is known for its efficiency and high performance in predicting high-dimensional data. However, due to the complex model structure, it generally suffers from the lack of methods for interpreting the meaning and significance of input variables. Among the three machine learning methods adopted in this paper, logistic regression yields more interpretable results due to the relatively simple relationship between inputs and outputs. The models developed in this paper exhibited good performance in that they reliably estimate the IRI given the pavement distress types and their extents and severities. However, the results showed that it is rather difficult to make such a conclusion for IRI relationships with any individual distress type. In this section of the paper, logistic regression was applied to measure the contribution of each distress type to surface roughness (IRI). Specifically, a binary logistic model was used in a bid to simplify the explanatory factors (levels of the pavement distresses). These were used to estimate the probability that the IRI is smaller or greater than a certain level given the levels of the pavement distresses. We herein present the results of two trained binary classifiers. The first sample classifier assigns the pavement sections into two groups: IRI < 100 and IRI > 100; and the second assigned them into classes of IRI <=200 and IRI >200.

Figures 10 (a) and (b) present the estimated coefficients of all the distress and other explanatory variables in the two developed sample logistic classifiers described above. In the first classifier, the width of the edge alligator crack, wheel-path longitudinal crack and wheel-path alligator crack at the high-severity level were found to be associated with the highest level of importance, with positive signs. This means the increase of the three factors lead to a significantly higher probability that the IRI falls in the ">100 in/mi" group. In the second model, the crack depth and width of longitudinal spall at medium and high levels were found to be the most influential variables that is associated with higher roughness (IRI > 200 in/mi). In addition, rutting was found to be highly correlated with IRI value, as indicated by the results of the two classifier specifications. A higher rutting was found to be associated with higher probability that the pavement has a high roughness level. Also, for both classifiers, jointed portland cement pavements were generally found to be more likely have a higher IRI value compared to other pavement types, and asphalt pavement was found to be more likely associated with a lower IRI. This suggests that for a given combination of pavement distresses, a higher IRI is exhibited by a rigid pavement compared to a flexible pavement. Further, where the second classifier was used for the analysis, it was found that Interstate pavements are influential: for a given combination of pavement distresses types, pavements on Intertate highways were found to exhibit lower surface roughness compared to their non-Interstate counterparts.

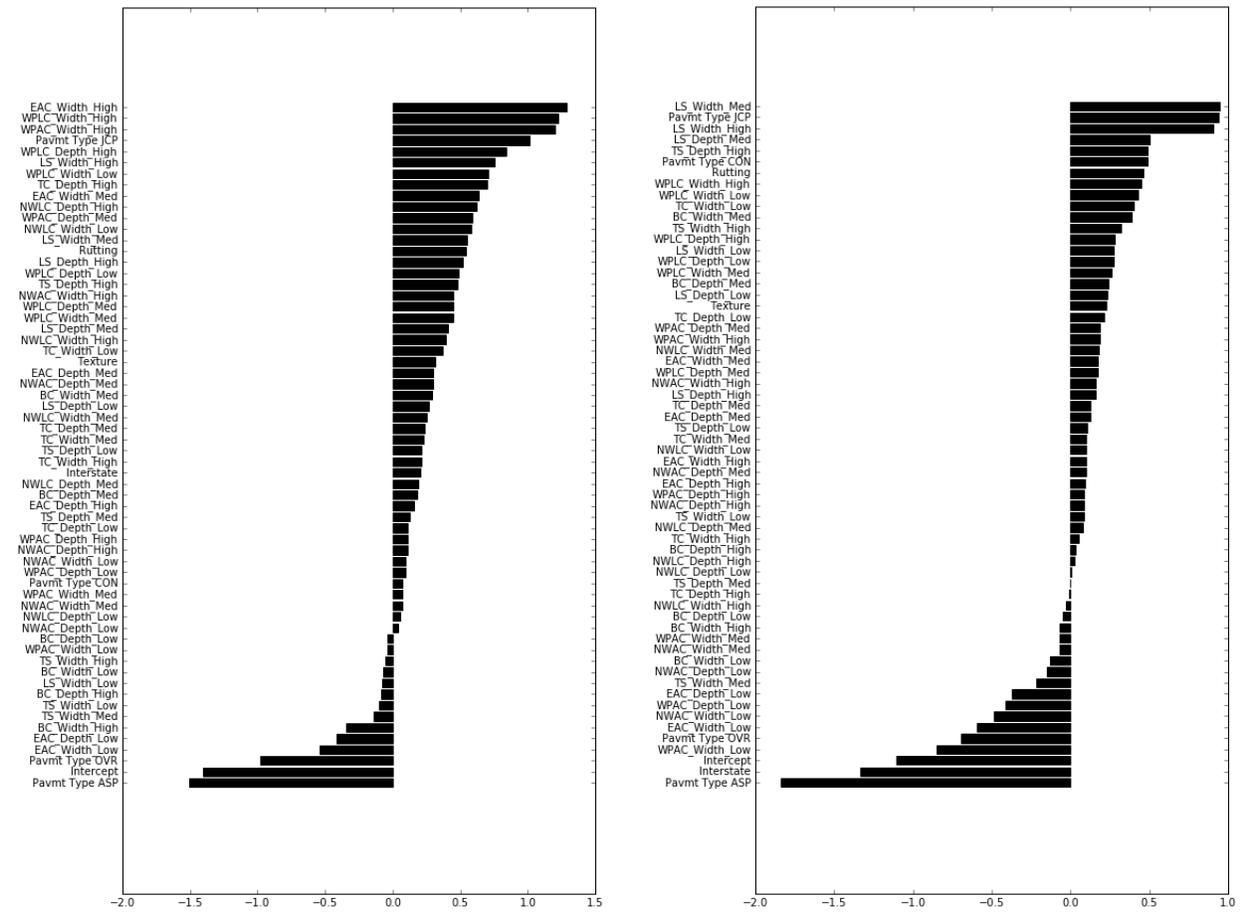

(a) Classifier IRI ≤ 100 vs. IRI > 100    (b) Classifier IRI ≤ 200 vs. IRI >200

Figure 9. Relative Importance of Distress Parameters (Two Sample Binary Logistic Classifiers)

**CONCLUDING REMARKS**
It is a truism that higher levels of distresses are generally associated with higher roughness, it is also true that a given roughness level can correspond to several possible combinations of distress types, extent, and severity. For this reason, the paper had sought to characterize numerically, the strength of this relationship, and to identify the pavement-related factors that influence this relationship. The paper developed models that can be used to estimate the surface roughness (IRI) of a pavement segment given the distribution of the various pavement distresses exhibited on that segment. The paper used machine learning methods due to the rather high dimensionality of the data. The prediction accuracy of the models was improved significantly when the data were aggregated from 0.0005 miles to 0.1 miles per section and when extremely large IRI values were identified as outliers and subsequently excluded from the analysis. Of the three machine learning techniques, the SVM classifier (with polynomial Kernel) yielded generally higher accuracy: the prediction accuracy is 0.86 (for IRI classified into a range of 30 in/mi groups), and 0.96 (for IRI classified into a range of 50 in/mi groups).

The paper showed that it is feasible to estimate the surface roughness of a given pavement based on the types of existing distressses and their occurrence attributes (extents and severities). The results also show that the pavement type and functional class infleunce the strength of this relationship. For a given combination of pavement distress types and their attrbutes, a higher IRI is exhibited by a rigid pavement compared to a flexible pavement. Further, it was found that for a given combination of pavement distresses types and their attributes, pavements on high class highways (Interstates) were found to exhibit lower surface roughness compared to relatively lower class highways (non-Interstates).

There are several avenues for future work. Lagged specifications of the model could be investigated. For example, overall performance (IRI) in a given year may be a function not of the distresses in that year but of the previous year. For example, increased cracking may not immediately result in increased surface roughness; however, over an extended period, cracks allow water to infiltrate the pavement layers (Ridgeway, 1976) and weakens the subbase or subgrade, and cause increased roughness. The time lag between crack manifestation and a bumpy surface ride may range from a few months to several years. The second part of this paper, where pavement distress types and their occurrence parameters (extent and severity) at a pavement section are estimated based on the section IRI, presents opportunities for future research. Specifically, future research could use advanced techniques to establish the expected distributions of the distress types and their occurrence variables for a given IRI value and how this probability is influenced by pavement and environmental attributes including the pavement class, surface material type, traffic levels, and climate. Other distributions besides Gaussian and other techniques besides regression, could be investigated to assess how well they fit the observed data, and to predict the probability that a specific pavement of known IRI has a certain type and severity of any given type of distress. In this paper, an exploratory analysis is done for cracking. Future work could examine other distresses besides cracking.

# REFERENCES

Adey, B.T., 2017. Process to enable the automation of road asset management, 2nd International Symposium on Infrastructure Asset (SIAM 2), June 29-30, 2017, ETH, Zurich.

Alinizzi, M., Chen, S., Labi, S., Kandil, A. (2018). A methodology to account for one-way infrastructure interdependency in preservation activity scheduling. Computer-Aided Civil and Infrastructure Engineering, 33(11), 905-925.

Al-Mansour, A., Alawal, R., 2006. Correlation of visual inspection and roughness measurement in pavement condition evaluation. Rep. 424/28, College of Engineering, King Saud Univ., Riyadh, Saudi Arabia.

Al-Masaeid, H.R., Al-Sharaf, J.S., Al-Suleiman, T.I., 1998. Effect of road roughness and pavement condition on traffic speed, REAAA Journal 11, 8-14.

Archondo-Callao, R. S., Faiz, A., 1994. Estimating vehicle operating costs, Tech. Paper No. 234, International Bank for Reconstruction and Development, World Bank, Washington, DC.

Barrette, T.P., 2011. Comparison of PASER and PCI pavement distress indices. M.S. thesis, Department of Civil and Environmental Engineering, Michigan Technological University, Houghton, MI.

Chandra, S., Sekhar, C.R., Bharti, A.K., Kangadurai, B., 2012. Relationship between pavement roughness and distress parameters for Indian highways, Journal of Transportation Engineering 139(5), 467-475.

Cheetham, A., Christison, T.J., 1981. The development of RCI prediction models for primary highways in the province of Alberta, Department of Transportation, Alberta, Canada.

Chan, W.T., Fwa, T.F., Hoque, K.H., 2001. Constraint handling methods in pavement maintenance programming, Transportation Research Part C 9, 175-190.

Friedman, J., Hastie, T., Tibshirani, R., 2001. The elements of statistical learning, Series in statistics, Springer, Berlin.

Gallego, C., Comendador, V., Nieto, F., Imaz, G., Valdésa, R., 2018. Analysis of air traffic control operational impact on aircraft vertical profiles supported by machine learning, Transp Res C,

Gao, L., Zhang, Z., 2008. Robust optimization for managing pavement maintenance and rehabilitation. Transportation Research Record 2084, 55-61.

Gillespie, T.D., 1980. Calibration of response-type road roughness measuring systems.

Greene, S., Akbarian, M., Ulm, F., Gregory, J., 2013. Pavement roughness and fuel consumption, MIT Concrete Sustainability Hub, Cambridge, MA.

Gulen, S., Woods, R., Weaver, J., Anderson, V.L., 1994. Correlation of present serviceability ratings with international roughness index. Transportation Research Record 1435, 27–36.

Hall, K.T, and Crovetti J.A., 2007. Effects of subsurface drainage on pavement performance, NCHRP Report 583, Transportation Research Board, Washington, D.C.

Hosmer, D., 2013. Applied logistic regression. Wiley, Hoboken, NJ.

James, G., Witten, D., Hastie, T., Tibshirani, R., 2013. An introduction to statistical learning Vol. 6, Springer, New York.

Jeong, H., Kim, H., Kim, K., Kim, H., 2017. Prediction of flexible pavement deterioration in relation to climate change Using Fuzzy Logic, Journal of Infrastructure Systems 23(4).

Karan, M. A., Haas, R., Kher, R., 1976. Effects of pavement roughness on vehicle speeds. Transportation Research Record 602, 122–127.

Knerr, S., Personnaz, L., Dreyfus, G., 1990. Single-layer learning revisited: a stepwise procedure for building and training a neural network. Neurocomputing 41-50. Springer, Berlin, Heidelberg.

Koutsopoulos, I., Kapotis, V.I., Downey, A.B., 1994. Improved methods for classification of pavement distress images, Transportation Research Part C 6816) 58–69.

Lasisi, A., Attoh-Okine, N., 2018. Principal components analysis and track quality index: A machine learning approach, Transportation Research Part C, 91(1), 230–248.

Lu, P., Tolliver, D., 2012. Pavement treatment short-term effectiveness in IRI change using long-term pavement program data. Journal of Transportation Engineering 138(11), 1297–302.

Lucas, J., Viano, A., 1979. Systematic measurement of evenness on the road network: high output longitudinal profile analyzer. French Bridge and Pavement Laboratories, Rep 101.

Mactutis, J., Alavi, S., Ott, W., 2000. Investigation of relationship between roughness and pavement surface distress based on WesTrack project. Transportation Research Record 1699 107–113.

Markovic, N., Milinkovic, S., Tikhonov, K.S., Schonfeld, P., 2015. Analyzing passenger train arrival delays with support vector regression, Transportation Res. Part C 56 (2015) 251–262.

Mariani, M.C., Bianchini, A., Bandini, P., 2012. Normalized truncated Levy walk applied to flexible pavement performance, Transportation Research Part C 24 (2012) 1–8.

McGhee, K. (2004). Automated pavement distress collection techniques, NCHRP Synthesis of Practice 334, Transportation Research Board.

Molenaar, A., Sweere, G.T., 1981. Road roughness: its evaluation and effect on riding comfort and pavement life, Transportation Research Record 836, 41–49.

Park, K., Thomas, N.E., Wayne Lee, K., 2007. Applicability of the international roughness index as a predictor of asphalt pavement condition, J. of Transportation Eng. 133(12), 706–709.

Parsley, L.L., Robinson, R., 1982. The TRRL road investment model for developing countries (RTIM2), No. LR 1057 Monograph.

Prasad, J. R., Kanuganti, S., Bhanegaonkar, P.N., Sarkar, A.K., Arkatkar, S., 2013. Development of relationship between Roughness (IRI) and visible surface distresses: A Study on PMGSY Roads. Procedia - Social and Behavioral Sciences,104, 322-331.

Ridgeway, H.H., 1976. Infiltration of water through the pavement surface, Transportation Research Record 616.

Sayers, M.W., 1986. Guidelines for conducting and calibrating road roughness measurements.

Sharif Tehrani, S., Cowe Falls, L., Mesher, D., 2017. Effects of pavement condition on roadway safety in the province of Alberta, Transportation Safety & Security 9(3), 259-272.

Vaillancourt, M., Houy, L., Perraton, D., Breysse, D., 2014. Variability of subgrade soil rigidity and its effects on the roughness of flexible pavements, Materials & Structures 48(11), 3527–3536.

Van Til, C.J., 1972. Evaluation of AASHO interim guides for design of pavement structures. Highway Research Board.

Wang, K. (2011). Elements of automated survey of pavements and a 3D methodology, Journal of Modern Transportation 19(1), 51–57.

Wang, K., and Gong, W. (2005). Real-time automated survey system of pavement cracking in parallel environment. J. Infrastructure Systems 11(3), 154–164.

Wang, T., Harvey, J., Lea, J., Kim, C., 2014. Impact of pavement roughness on vehicle free-flow speed, Journal of Transportation Engineering 140(9).

Wilson, S.P., Harris, N.K., Obrien, E.J., 2006. The use of Bayesian statistics to predict patterns of spatial repeatability, Transportation Research Part C 14, 303–315.

Yuan, F., Cheu, R.L., 2003. Incident detection using support vector machines, Transportation Research Part C 11, 309–328.

Zaabar, I., Chatti, K., 2010. Calibration of HDM-4 models for estimating the effect of pavement roughness on fuel consumption for U. S. conditions. Transportation Research Record 2155, 105-116.

Zalama, E., Gòmez-Garcià-Bermejo, J., Medina, R., Llamas, J. (2014). Road crack detection using visual features extracted by Gabor filters. Comp.-Aided Civ. Infra. Eng., 29(5), 342–358.

Zhou, G. Wang, L., 2012 Co-location decision tree for enhancing decision-making of pavement maintenance and rehabilitation, Transportation Research Part C 21 287–305.

Chen, S., Tang, Z., Zhou, H., Cheng, J. (2019). Extracting topographic data from online sources to generate a digital elevation model for highway preliminary geometric design. Journal of Transportation Engineering, Part A: Systems, 145(4), 04019003.

Chen, S. (2019). Safety implications of roadway design and management: new evidence and insights in the traditional and emerging (autonomous vehicle) operating environments. Ph.D. dissertation, Purdue University, West Lafayette.

Chen, S., Saeed, T. U., Alqadhi, S.D., Labi, S. (2017). Safety impacts of pavement surface roughness at two-lane and multi-lane highways: accounting for heterogeneity and seemingly unrelated correlation across crash severities. Transportmetrica A: Transport Science, 15(1), 18-33.

Chen, S., Saeed, T.U., Alinizzi, M., Lavrenz, S., Labi, S. (2019). Safety sensitivity to roadway characteristics: A comparison across highway classes. Accident Analysis & Prevention, 123, 39-50.

Tang, Z., Chen, S., Cheng, J., Ghahari, S. A., Labi, S. (2018). Highway design and safety consequences: A case study of interstate highway vertical grades. Journal of Advanced Transportation, 2018. Article ID 1492614

Chen, S., Saeed, T.U., Labi, S. (2017). Impact of road-surface condition on rural highway safety: A multivariate random parameters negative binomial approach. Analytic Methods in Accident Research, 16, 75-89.

Labi, S., Chen, S., Preckel, P.V., Qiao, Y., Woldemariam, W. (2017). Rural two-lane highway shoulder and lane width policy evaluation using multiobjective optimization. Transportmetrica A: Transp. Sci. 13(7), 631-656.

Chen, S., Leng, Y., Labi, S. (2020). A deep learning algorithm for simulating autonomous driving considering prior knowledge and temporal information. Computer‐Aided Civil and Infrastructure Engineering, 35(4), 305-321.

Chen, T., Sze, N.N., Chen, S., Labi, S. (2020). Urban road space allocation incorporating the safety and construction cost impacts of lane and footpath widths. Journal of safety research, 75, 222-232.

Ghahari, S., Alabi, B., Alinizzi, M., Chen, S., Labi, S. (2019). Examining the relationship between infrastructure investment and performance using state-level data. Journal of Infrastructure Systems, 25(4), 04019026.

Ghahari, S., Chen, S., Labi, S. (2021). A Nonparametric Efficiency Methodology for Comparative Assessment of Infrastructure Agency Performance. Transportation Engineering, 6, 100092.

Chen, T., Sze, N.N., Chen, S., Labi, S., Zeng, Q. (2021). Analysing the main and interaction effects of commercial vehicle mix and roadway attributes on crash rates using a Bayesian random-parameter Tobit model. Accident Analysis & Prevention, 154, 106089.

Kaseko, M.S., Ritchie, S., 1993. A neural network-based methodology for pavement crack detection and classification, Transpn. Res.-C. 1(4), 275–291.

Dougherty, M., 1995. A review of neural networks applied to transport, Trans. Res.C. 3(4), 247–260.

Domingos, P., Pazzani, M., 1997. On the optimality of the simple Bayesian classifier under zero-one loss. Machine Learning 29, 103–130.

Owusu-Ababio, S., 1998. Effect of neural network topology on flexible pavement cracking prediction. Computer-aided Civil & Infrastructure Engineering 13(5), 349–355.

Archilla, A.R., Madanat, S., 2000. Development of a pavement rutting model from experimental data. Journal of Transportation Engineering 126(4), 291-299.

Gharaibeh, N.G., Zou, Y., Saliminejad, S., 2009. Assessing the agreement among pavement condition indexes. J. Transport. Eng. 136 (8), 765–772.

Haider, S., Dwaikat, M., 2011. Estimating optimum timing for preventive maintenance treatment to mitigate pavement roughness. Transportation Research Record 2235, 43–53.

Bai, Q., Ahmed, A., Li, Z., Labi, S., 2014. A hybrid Pareto frontier generation method for trade-off analysis in transportation asset management, Computer-aided Civil & Infrastructure Engineering 30(3), 163–180.

Meegoda, J. N., Gao, S., 2014. Roughness progression model for asphalt pavements using long-term pavement performance data. Journal of Transportation Engineering 140(8), 04014037.